\documentclass[12pt,preprint]{aastex}

\def\simless{\mathbin{\lower 3pt\hbox
     {$\rlap{\raise 5pt\hbox{$\char'074$}}\mathchar"7218$}}}   %< or of order
\def\simmore{\mathbin{\lower 3pt\hbox
     {$\rlap{\raise 5pt\hbox{$\char'076$}}\mathchar"7218$}}}   %> or of order
\def\av#1{\langle #1\rangle}
\def\hide#1{}

\received{}
\accepted{}
\slugcomment{ApJ Main Journal MS\# 54035; submitted 2001 May 11; 
revised version 2001 July 11}

\lefthead{M. van der Klis}
\righthead{Parallel Lines in kHz QPOs}

\begin{document}

\title{A Possible Explanation for the ``Parallel Tracks''
Phenomenon in Low-Mass X-Ray Binaries}

\author{M. van der Klis\altaffilmark{1}
}

\altaffiltext{1}{Astronomical Institute ``Anton Pannekoek'',
       University of Amsterdam and Center for High-Energy Astrophysics,
       Kruislaan 403, NL-1098 SJ Amsterdam, the Netherlands}

\begin{abstract}
An explanation is proposed for the observation that in low-mass X-ray
binaries the correlation between most observable X-ray spectral and
timing parameters on the one hand, and X-ray luminosity on the other,
while generally good in a given source on a time scale of hours, is
absent both on longer time scales and between sources. This
phenomenon, particularly evident in kHz QPO sources, leads to parallel
tracks in plots of such parameters vs. luminosity. It is pointed out
that where previously proposed explanations require at least two
time-variable independent parameters, such as accretion rate through
the disk and through a more radial inflow, just one independent
variable is in fact sufficient, provided that the systemic response to
time variations in this variable has both a prompt and a time-averaged
component. A specific scenario is explored in which most observable
spectral and timing parameters to first order depend on disk accretion
rate normalized by its own long-term average rather than on any
individual accretion rate (luminosity, on the contrary, just depends
on the total accretion rate). This provides a way in which parameters
can be uncorrelated to accretion rate, yet vary in response to
variations in accretion rate.  Numerical simulations are presented of
such a model describing the relation between kHz QPO frequency and
X-ray luminosity, which observationally is characterized by a striking
pattern of parallel tracks in the frequency vs. luminosity plane both
in individual sources and across sources.  The model turns out to
reproduce the observations remarkably well. Physical interpretations
are suggested that would produce such a scenario; particularly
promising seems an interpretation involving a radial inflow with a
rate that derives through a time-averaging process from the disk
accretion rate, and an inner disk radius that depends on the balance
between the accretion through the disk and the total luminosity.  The
consequences of this idea for our understanding of states and tracks
in LMXBs are discussed, and the applicability of the idea to
black-hole candidates, where the observational situation is more
complex, is briefly addressed.
\end{abstract}


\section{Introduction}

The relation between the X-ray spectral/timing states of low-mass
X-ray binaries (LMXBs) and their X-ray luminosity is puzzling. While
on short time scales (typically, hours) X-ray spectral and timing
parameters vary (stochastically) in a well-correlated fashion, on
longer time scales (typically, days) the X-ray luminosity,
particularly in the low-energy band ($<6$\,keV) appears to vary
largely independently from the other parameters. This is seen in the
near-Eddington neutron-star Z sources (e.g., Hasinger et al. 1990,
Kuulkers et al. 1996, Wijnands et al. 1998, Jonker et al. 2000) as
well as in the less luminous neutron-star atoll sources (e.g., van der
Klis et al. 1990, Ford et al. 1997). Observations that could be
interpreted in terms of a similar phenomenon in black-hole candidates
(e.g., M\'endez and van der Klis 1997, Rutledge et al. 1999) have
recently been provided with solid empirical underpinnings by a study
of the black-hole candidate XTE J1550$-$564 (Homan et al. 2001).

In all these sources, timing properties, in particular the
characteristic frequencies of QPO and noise components correlate much
better to each other (e.g., Ford and van der Klis 1998), and to
measures of X-ray spectral shape such as position in the tracks LMXBs
tend to trace out in an X-ray color-color diagram (e.g. M\'endez and
van der Klis 1999), X-ray colors (M\'endez et al. 1999; particularly
at higher energy), and sometimes parameters obtained from X-ray
spectral fits (Kaaret et al. 1998), than to luminosity (see van der
Klis 2000), although weak dependencies of timing and spectral
properties on long-term luminosity variations are observed (Kuulkers
et al. 1996, Wijnands et al. 1997). (For example, spectral shape in
the $<$6\,keV band is usually somewhat affected by the longer time
scale variations, spectral hardness in this band being positively
correlated with flux).  The time scale on which the correlated
short-term variability takes place is hours to days in the Z sources
and in atoll sources in luminous states, but it is sometimes longer in
black-hole candidates and in atoll sources in low-luminosity states.

So, while on short time scales all properties including luminosity
vary in a well-correlated fashion, on longer time scales X-ray
luminosity is the odd one out in an otherwise still rather
well-correlated set of spectral and timing parameters.  When for a
given source a plot is made of any spectral or timing parameter
vs. X-ray flux, then, due to this phenomenon, the result tends to be a
series of parallel tracks, each individual track reflecting the
short-term correlated variations at a given epoch, and the offsets
between the tracks resulting from the longer-term luminosity
variations, combined with observational windowing.  Perhaps the most
direct measurements of this parallel-track phenomenon have been made
using the frequencies of the kilohertz quasi-periodic oscillations
(kHz QPOs) in the Z and atoll sources. For this reason, most of the
remainder of this paper focuses on the parallel track phenomenon as
seen in kHz QPO sources (Fig.\,1); I return to the general case in
\S\ref{sect:disc}.

\subsection{Parallel tracks and kilohertz QPOs}

Kilohertz QPOs are observed in the X-ray flux of some twenty Z and atoll
sources and are generally interpreted as being due to the motion of
matter within a few stellar radii of the low-magnetic-field neutron
star in these systems (van der Klis 2000 for a review). Most models
involve tight orbital motion of matter around the star at a preferred
radius $r$ in the inner accretion disk, where $r=12-25$\,km. Because
of this proximity of their site of origin to the compact object, kHz
QPOs can potentially be used to constrain theories of strong gravity
and dense matter.

A discussion of the physics which might determine $r$ was given by
Miller et al. (1998) (see also Miller and Lamb 1993, 1996). In their
description, $r$ is the inner radius of the Keplerian disk, which as
long as it is larger than the innermost stable orbit from general
relativity, is set by radiation drag on the orbiting matter. For a
constant mass inflow, the radius $r$ increases when the X-ray flux
impinging upon the inner edge of the disk increases; for a constant
radiation field, $r$ decreases when the amount of matter accreting in
the disk increases. When both mass flow and radiation increase, it is
not obvious what will be the effect on $r$, but according to Miller et
al. (1998) if all radiation is due to accretion through the disk at a
rate $\dot M$, then $r$ will decrease when the accretion rate
increases. Under certain simplifying assumptions, $r\propto\dot
M^{3/2}$. Other proposed kHz QPO models involving the frequency of
orbital motion (e.g., Stella and Vietri 1999, Titarchuk et al. 1999,
Cui et al. 1998, Cui 2000, Campana 2000, Psaltis and Norman 2000)
either adopt the Miller et al. (1998) mechanism, or rely on a
magnetosphere to set the inner disk radius.

The dependence of kHz QPO frequency $\nu$ on X-ray flux $F_x$ shows
the above-described parallel track pattern. Whereas in a given source
on short time scales (less than several hrs) a good correlation is
observed between $\nu$ and $F_x$, on longer time scales (more than a
day or so) the correlation breaks down. On those longer time scales,
different flux levels can correspond to approximately similar
frequency ranges. Observational windowing has so far prevented a
careful study of what happens on intermediate time scales. In a
frequency vs. flux plot of a series of observations of a given kHz QPO
source, each lasting several hours and separated by intervals of days
or longer, a series of parallel tracks is observed, each track
corresponding to the $\nu$, $F_x$ relation of a single observation.

These parallel tracks are perhaps most clearly seen in the transient
atoll source 4U\,1608$-$52. In Fig.\,1{\it a}; M\'endez et al. 1999)
the parallel tracks are obvious; they are steeply inclined,
approximately straight lines. Note how there seems to be a systematic
trend in the steepness of these lines: at lower count rate they tend
to be steeper. The range in flux over which in a given source
identical frequencies are observed can be as large as a factor of
3. Note that there appears to be a tendency for the lines at the
highest flux to cover a somewhat higher frequency range, and vv.

A similar and perhaps related, but more extreme picture is presented
by observations of different sources: the same range of QPO
frequencies is observed in sources that differ in X-ray luminosity
$L_x$ (defined here as $4\pi d^2F_x$ where $d$ is the distance) by up
to more than two orders of magnitude (van der Klis 1997, Ford et
al. 2000). In a plot of frequency vs. luminosity covering the ensemble
of kHz QPO sources this once again leads to a series of parallel
tracks, each ``track'' now corresponding to a different source
(Fig.\,1{\it b}; Ford et al. 2000). Note, that these tracks are
approximately parallel in a plot of frequency vs. the {\it logarithm}
of $L_x$, so the same trend of steeper lines at lower luminosity as
observed in 4U\,1608$-$52 clearly also applies across the ensemble of
sources. (In a semi-logarithmic version of the plot in Fig.\,1{\it a},
the lines are also more parallel). In both figures it is as if
frequency depends not on $L_x$, but on the percentage by which $L_x$
deviates from its average (cf. van der Klis 1999).

\subsection{Proposed explanations}

The question in both cases illustrated in Fig.\,1 of course is: if at
some epoch in some source frequency depends on luminosity according to
some particular relation, then why at some other epoch or in some
other source is there a very different frequency-luminosity relation,
where very similar frequencies are attained at very different
luminosities? Any explanation that succeeds in obtaining the {\it
same} frequency at quite {\it different} luminosity would at first
sight seem to defy itself, as it would at the same time remove the
observed strong dependence of frequency on luminosity at any given
epoch. As explained below, up to now the answer typically has been
that at least two independently varying free parameters must
characterize the problem.

The phenomenology clearly suggests that a single quantity, variable on
time scales of less than a few hours, dominates kHz QPO frequency and
most other spectral and timing parameters, but not the observed X-ray
luminosity.  Usually, this parameter has been inferred to be the mass
accretion rate $\dot M$, sometimes parametrized by the curve length
$S$ measured along the track the source traces out on these time
scales in the X-ray color-color diagram (\S1; Hasinger and van der
Klis 1989, van der Klis 1995 for a review). A problem with this
interpretation is that the X-ray luminosity, whose energy ultimately
derives from this mass accretion, is on longer time scales not then
well-correlated to inferred $\dot M$.  Another parameter, varying on
time scales of more than a day or so, is therefore required to explain
the slow luminosity variations that do not affect the kHz QPO
frequency, i.e., the horizontal offset between the parallel lines in a
given kHz QPO source (Fig.\,1{\it a}). Slow variations in the
accretion geometry leading to changes in X-ray beaming (e.g., van der
Klis 1995) or in the ratio between matter accreting in the disk and
through some other channel such as a more radial inflow (e.g.,
Wijnands et al. 1996, Kaaret et al. 1998) energy loss modes in
unobservable spectral bands (e.g., van der Klis et al. 1995) and
mechanical energy in jets (e.g., M\'endez et al. 1999), have all been
suggested to provide this additional parameter, but in the absence of
a specific mechanism to make these parameters vary independently from
inferred $\dot M$, the true origin of the slow luminosity variations
has been difficult to pin down (see Ford et al. 2000 for a discussion
of these various possibilities).

Yet another parameter, different from source to source, is required in
this picture to explain the parallel lines spanning two orders of
magnitude in $L_x$ accross the ensemble of kHz QPO sources
(Fig.\,1{\it b}). It has been proposed that this parameter is the
neutron-star magnetic-field-strength, which should then be correlated
to average source luminosity rather tightly (White and Zhang 1997).
Systematic differences between sources with respect to one of the
slowly varying parameters mentioned above could perhaps explain
Fig.\,1{\it b}, but it has remained unclear which parameter could
fulfill all requirements (e.g., Ford et al. 2000). All proposals to
date effectively require the existence of at least two independent
time-variable parameters, and no compelling description was given of a
mechanism that could make such a second time-variable parameter vary
on its own accord independently of $\dot M$.

It is the purpose of this paper to point out that the ensemble of
observations described above can in fact be explained in terms of just
a single time-variable independent parameter, varying on time scales
of hours. This can be accomplished by exploiting the observed
difference in time scale between correlated and uncorrelated
behavior. The crucial requirement is that there exists both a prompt
and a time-averaged component in the systemic response to the
variations in the governing parameter.  The basic class of scenarios
that could accomplish this is outlined in \S2, and a specific proposal
is made there, namely that the governing parameter is the mass
accretion rate through the disk. In \S3 it is quantitatively
investigated to what extent such a model could explain the
observational facts presented above.  Numerical simulations of the
predicted luminosity, kHz QPO frequency relation in a representative
example of such a scenario are described there. In \S4 I discuss the
results, and the extent to which the scenario might be applicable to
the parallel-track phenomena observed in neutron-star and black-hole
low-mass X-ray binaries in general, as well as some of the the wider
implications of the proposed explanation for the interpretation of
states and tracks in LMXBs.

\section{Scenario}\label{sect:scen}

If the kHz QPO frequency $\nu$ depends on some parameter $x(t)$ which
varies on time scales of less than several hours, and the X-ray
luminosity $L_x$ depends partly on $x(t)$ and for another part on some
running average of $x$ such as $\av{x}(t) =
\int_{-\infty}^{+\infty}{w(t'-t)x(t')dt'}$, where $w(t)$ is a weight
function with integral 1 that is non-zero on interval $[t_1,t_2]$
spanning between several hours and days, then this could explain the
parallel tracks in a given kHz QPO source. The correlation between
$\nu$ and $L_x$ on time scales of hours would arise because both
quantities depend on $x(t)$, the decorrelation on longer time scales
because the slowly varying $\av{x}(t)$ affects $L_x$, but not $\nu$.

However, such a description would for plausible prescriptions for
$\av{x}$ have trouble to explain that a similar range of frequencies
is seen in 4U\,1608$-$52 for fluxes different by a factor of 3, and
could certainly not explain that the same frequencies are observed
over several orders of magnitude in luminosity across sources. For
$L_x$ to be very different, $x$, and hence $\nu$ would have to be very
different, contrary to what is observed. Also, in general the lines
would not be parallel in $\nu$ vs. $\log L_x$ space.

With an additional twist, such a scheme {\it can} explain both
parallel-track phenomena. Suppose, for example, that $\nu$ is
proportional to $\eta\equiv x/\av{x}$, but $L_x$ is proportional to
$x$ only. Then, when $x$ increases, $\nu$ and $L_x$ immediately
increase in correlation with it, but if $x$ remains the same for a
while, then on a time scale of more than several hours $\nu$ would
gradually drift to the value it has when $\eta=x/\av{x}=1$. In such a
picture a 100 times more luminous source can still show the same QPO
frequency as its weaker kin because not only $x$ but also $\av{x}$ is
approximately 100 times higher, so that $\eta=x/\av{x}$ is still
approximately the same. Note also that in such a scheme the same
fractional variation in $x$ tends to cause a similar absolute change
in $\nu$, so that the lines are parallel in $\nu$, $\log L_x$ space,
as observed. Several similar prescriptions for the dependence of $\nu$
and $L_x$ on $x$ and $\av{x}$ are possible that preserve the desirable
qualities of this description.

An attractive interpretation retaining the flavor of this basic idea
that would work well within the framework of a model where the inner
radius $r$ of the disk, and hence $\nu$, is determined by some balance
between accretion rate through the disk and radiative stresses (\S1.1),
would be one where $\nu$ depends on the ratio between instantaneous
accretion rate through the disk $\dot M_d(t)$ varying on short time
scales, and $L_x$:
$$\nu(t)\propto\left({\dot M_d(t)\over L_x(t)}
\right)^\beta,\qquad\qquad\ \ \ \ \beta>0\qquad\qquad\ \hbox{(1a)}$$
and the luminosity has both an immediate
response to $\dot M_d$ and a filtered one: 
$$L_x(t)\propto\dot M_d(t)+\alpha\av{\dot M_d}(t),\qquad\alpha>0
\qquad\qquad\hbox{(1b)}$$
So, in this model $x(t)$ is identified with the accretion rate $\dot
M_d(t)$ through the disk, $\nu$ depends on $x/(x+\alpha\av{x})$, and
$L_x$ on $x+\alpha\av{x}$. Note that $x/(x+\alpha\av{x}) =
(1+\alpha/\eta)^{-1}$ where $\eta\equiv\dot M_d/\av{\dot M_d}$, so
$\nu$ still only depends on the ratio $\eta$ as in the simpler
prescription above. For this reason, the model retains the basic
qualities of this simpler prescription. At epochs where $\eta<1$,
i.e., $\av{\dot M_d} > \dot M_d$ too much radiation is produced for
the amount of matter in the disk (set by $\dot M_d$) so $r$ is large
and hence $\nu$ low, and vv. If $\dot M_d$ remains the same for a
while, then $\av{\dot M_d}$ approaches $\dot M_d$, i.e.,
$\eta\approx1$, and $r$ and hence $\nu$ drift to their ``equilibrium''
value, no matter what the actual value of $\dot M_d$ is.  The lines
tend to be parallel in $\nu$, $\log L_x$ space as, depending only on
the extent to which $\av{\dot M_d}$ has converged to $\dot M_d$, the
same fractional change in $\dot M_d$ and hence $L_x$ tends to produce
the same linear change in $\nu$.  Note that no other independently
time-varying parameter than $\dot M_d$ is required now to explain the
parallel-tracks phenomenon.

The physical question is: what is the nature of the slowly varying
component in the response of luminosity to accretion rate, the term
$\alpha\av{\dot M_d}(t)$? Physical implementations could be imagined
using for example a mechanism for the storage and slow release of
energy in the neutron star, such as through deposition and burning of
nuclear fuel, energy storage in the crust or in the disk/star boundary
(Inogamov and Sunyaev 1999, Popham and Sunyaev 2001). Another
possibility would be energy storage in the neutron-star spin, with
energy feeding back into the disk by magnetic or gravitational
interaction (cf. Priedhorsky 1986).  Alternatively, one could think of
a mechanism where matter flowing in radially provides a second channel
of accretion whose $\dot M$ is derived from some running average over
$\dot M_d$, for example because more matter in the disk leads to more
matter blowing off it to feed the radial flow, or because disk and
radial flow are both fed by the same external flow.

Note from these examples that the filtered response can technically be
either causal (e.g., burning rate depends on amount of fuel accreted
in the past), or acausal (e.g., matter blown from disk reaches neutron
star before disk flow); the response may even be
symmetric\hide{check}. The basic feature of the model scenario
discussed here is that more accretion through the disk produces more
flux not only promptly, but also through some time-averaged response,
and (only) when this filtered response is at the level it would have
when the accretion were constant, do the effects on the QPO frequency
cancel.

\section{Simulations}\label{sect:simulations}

\begin{figure}[htb]
\epsscale{1.0}
\plotone{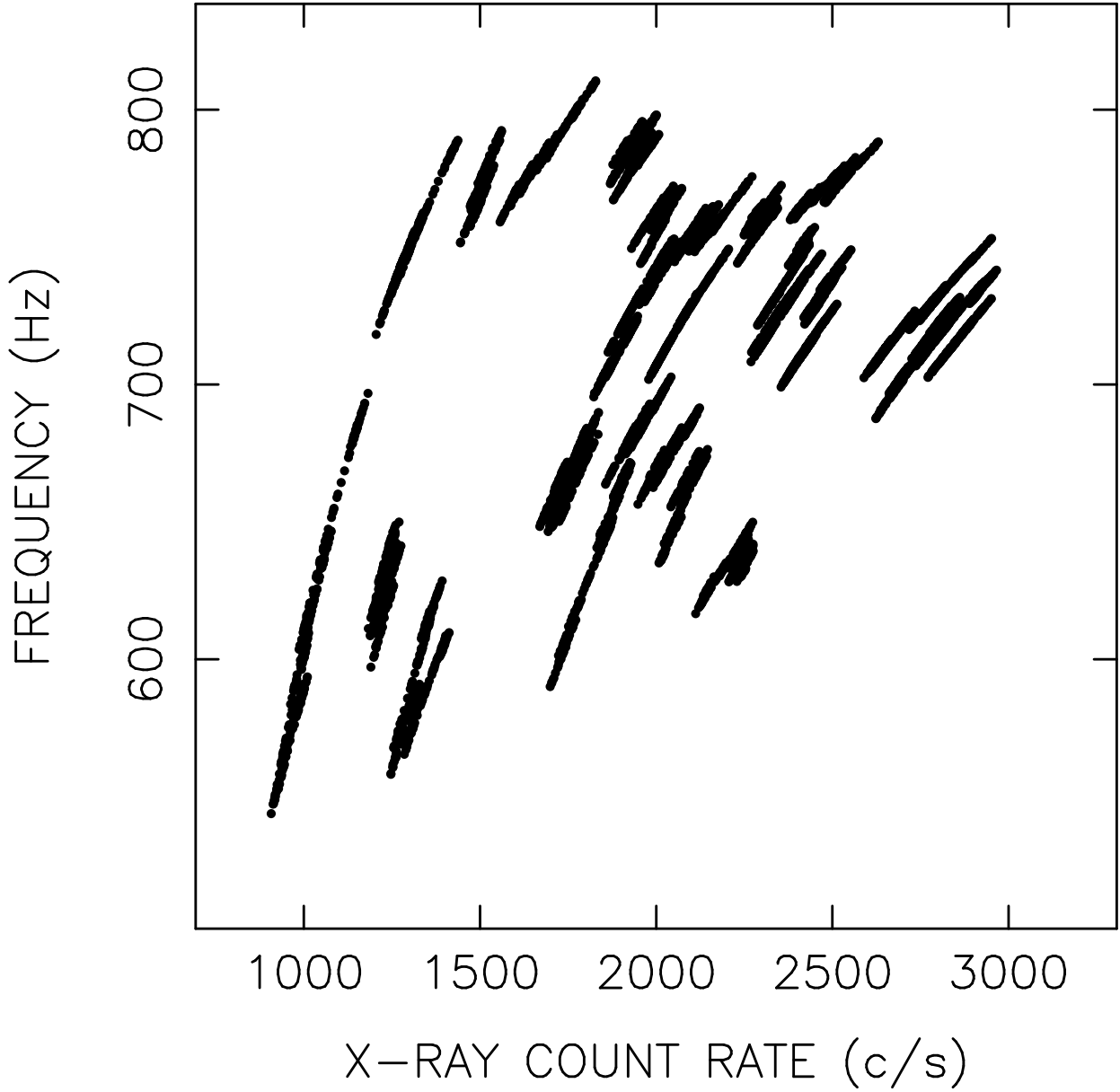}
\caption{Parallel lines in the frequency
vs. count rate diagram produced by the model described in
\S\ref{sect:simulations}. Data gaps of 16 hours were introduced in the
12-day simulated time series to mimick typical observational
windowing.
\label{fig:simulation1608}}
\end{figure}

For definiteness, assume the following description for the dependence
of frequency on accretion rate and X-ray luminosity:
$$\nu = \nu_0\left({\epsilon\dot M_d(t)c^2\over L_x(t)}\right)^\beta =
\nu_0\left({\dot M_d(t)\over \dot M_d(t) + \alpha\av{\dot
M_d}(t)}\right)^\beta \hbox{(2a)}$$
where the luminosity is simply $L_x(t)/c^2 = \epsilon(\dot M_d(t) +
\dot M_{rad}(t))$ with $\epsilon$ the accretion mass-to-energy
conversion efficiency, and where the radial accretion rate is related
to the disk accretion rate via $\dot M_{rad}(t) = \alpha\av{\dot
M_d}(t)$. So, in this description $\eta\equiv\dot M_d/\av{\dot
M_d}=\alpha\dot M_d/\dot M_{rad}$. Frequency $\nu$ only depends on
$\eta\propto\dot M_d/\dot M_{rad}$ and $L_x$ on $\dot M_d+\dot
M_{rad}$. The filtered response $\av{\dot M_d}$ is given by an
integral over the future disk accretion rate:
$$\av{\dot M_d}(t) = {1\over\tau} \int_{t}^{\infty}
{e^{-(t'-t)/\tau}\dot M_d(t')dt'}. \qquad\qquad\hbox{(2b)}$$
Results of numerical simulation of this description with a
radial-to-disk accretion ratio $\alpha$ of 1/3, filter time scale
$\tau=2\,\hbox{days}$ and $\beta=2$ are shown in
Fig.\,\ref{fig:simulation1608}, where a second-order red-noise signal
(random walk) was used for $\dot M_d(t)$, and 12 days of data were
plotted in intervals of 8\,hrs separated by gaps of 16\,hrs to
simulate typical observational windowing. Note that these choices for
$\tau$ and windowing time scales are illustrative; the scenario will
work for a variety of time scales.

Comparing Fig.\,2 to Fig.\,1{\it a}, clearly the result of this
simulation is rather similar to observation. Not only do the parallel
lines show up, also the steepening of the tracks towards lower
luminosity is reproduced, as well as the tendency for lines at higher
count rates to cover higher frequency ranges. As explained in \S2, the
steepening effect is a consequence of the fact that frequency depends
only on $\eta$, so that same fractional change in $\dot M_d$ tends to
produce the same linear change in $\nu$. The other effect arises
because the highest count rates tend to be attained at the highest
disk accretion rates $\dot M_d$. Necessarily, these tend to be
followed by intervals of lower $\dot M_d$, so that $\av{\dot M_d}$
tends to be lower than $\dot M_d$ at those epochs, i.e., $\eta>1$,
which explains why relatively higher frequencies often occur there. At
low count rates the opposite applies.

\begin{figure}[htb]
\plotone{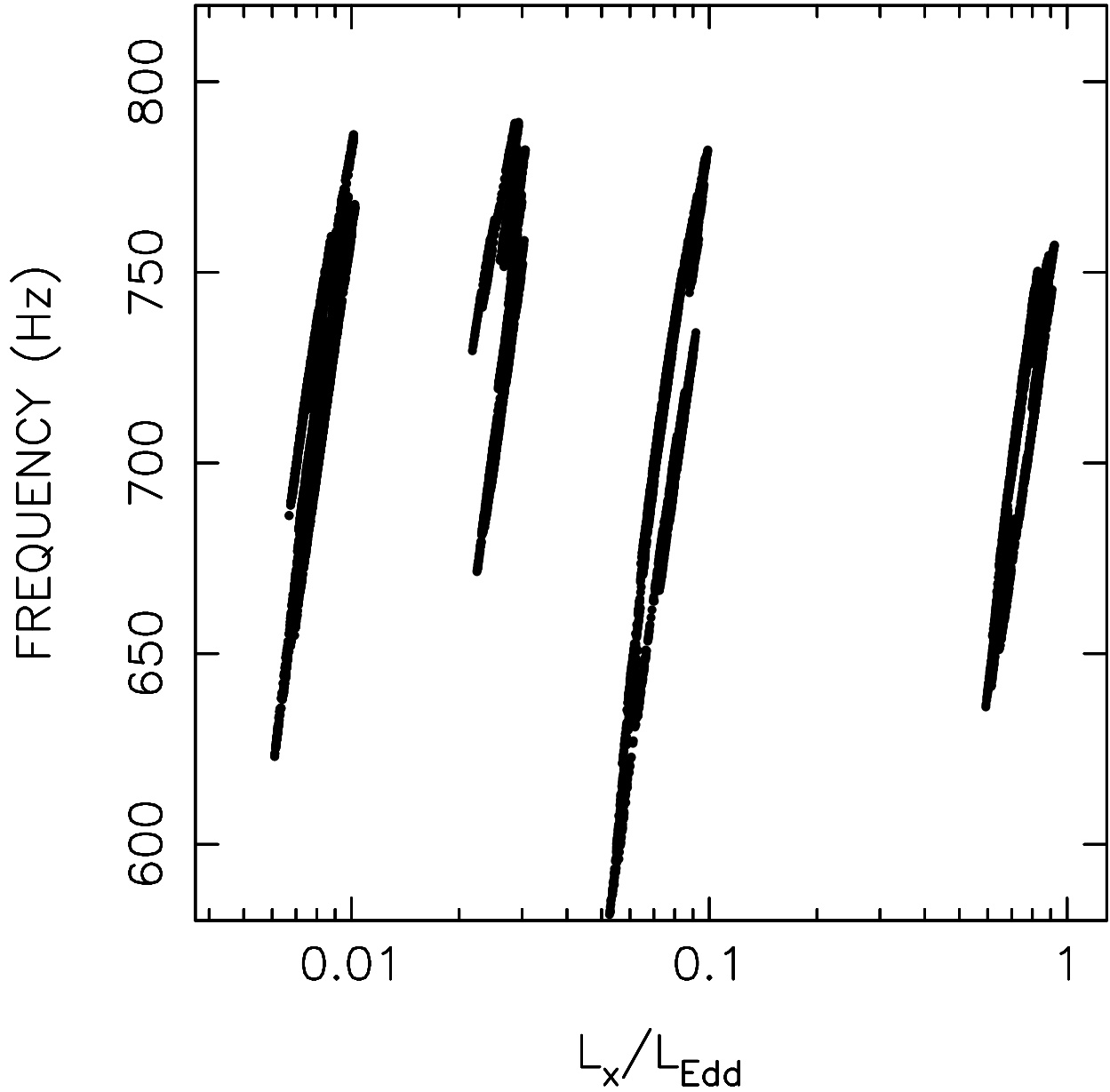}
\caption{Parallel lines in the frequency
vs. X-ray luminosity diagram for 4 different simulated sources
produced by the model described in \S\ref{sect:simulations}. Each
source was followed for 5 days, during which it varied randomly in
luminosity by a factor of 2.
\label{fig:simulationallsources}}
\end{figure}

Fig.\,\ref{fig:simulationallsources} shows the result of the
simulation of four sources differing in average $\dot M_d$ but showing
a similar fractional amplitude of $\dot M_d$ variations. Each source
was simulated for 5 days, during which it varied randomly in
luminosity by a factor of 2. Comparing Fig.\,3 to Fig.\,1{\it b},
again the result of the simulation is similar to what is observed. In
particular, the simulated tracks distributed over a factor $\sim$100
in $L_x$, are approximately parallel in $\nu$, $\log L_x$ space.

Very similar results were obtained from a simulation where the
filtered response was representative of one due to nuclear burning of
accreted material rather than of radial accretion. In that case,
$L_x/c^2 = \epsilon_{acc}\dot M_d(t) + \epsilon_{nuc}\av{\dot
M_d}(t)$ with $\epsilon_{acc}$ and $\epsilon_{nuc}$, respectively,
the accretion and the nuclear burning mass-to-energy conversion
efficiencies, typically 0.1 and 0.003, and $\av{\dot M_d}(t)$ now given
by a {\it past} integral over $\dot M_d(t)$. However, in order for the
observed steep dependencies of $\nu$ on $L_x$ to be reproduced in
those simulations, a large value of $\beta$ between 10 and 15 had to
be used. This is because of the small value of $\alpha =
\epsilon_{nuc}/\epsilon_{acc} = 0.03$ appropriate to that physical
example. The work of Miller and Lamb (1993, 1996) suggests that lower
values of $\beta$ such as the value of 2 used in the first example are
more appropriate.

The velocity of a radial flow is expected to be much higher than the
radial component of the motion of the matter accreting in the
disk. This is expressed in Eq.\,2b by making $\dot M_{rad}$ reflect
the {\it future} evolution of $\dot M_d$. Therefore, it might seem
surprising that the variations in $\dot M_{rad}$ would be a low-pass
filtered version of, i.e., slower than, those in $\dot M_d$. However,
note that the disk is expected to feed the radial flow over an annulus
covering a range of radii. The density fluctuations in the disk would
contribute to the radial flow for the full time they take to cross
this 'feeding annulus', i.e., for much longer than they take to cross
the inner disk edge and accrete in the end. Alternatively, the
short-term fluctuations in $\dot M_d$ could form at small radii in the
disk, within the feeding annulus. In both cases $\dot M_{rad}$ would
be a smoothed version of $\dot M_d$, which is all that is required to
make the scenario work.

\section{Discussion}\label{sect:disc}

\begin{figure}
\epsscale{1.0}
\plottwo{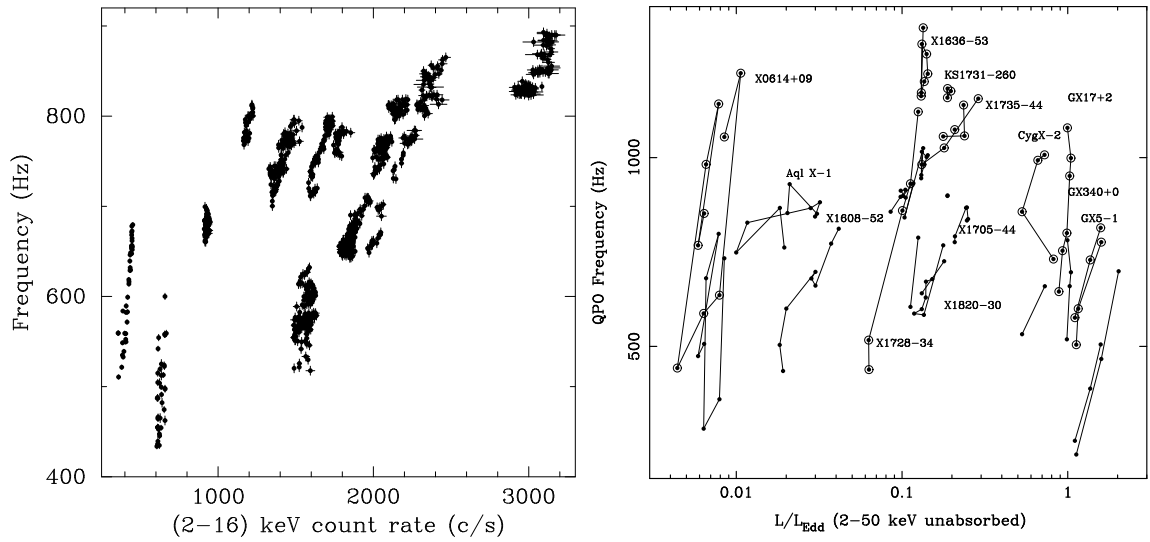}{f1b.eps}
\caption{a. ({\it left}):
Lower kHz QPO frequency vs. count rate in 4U\,1608$-$52 showing the
``parallel lines'' phenomenon as it occurs in a single source. After
M\'endez et al. (1999). b. ({\it right}): Kilohertz QPO frequencies
vs. X-ray luminosity in 13 different sources showing the parallel
lines phenomenon across sources. After Ford et al. (2000).
\label{fig:parlines1608}\label{fig:parlinesallsources}}
\end{figure}

The parallel tracks observed in kHz QPO sources in plots of QPO
frequency vs. logarithmic X-ray luminosity within a given source as
well as across sources (Fig.\,\ref{fig:parlines1608}) can be explained
in terms of just one time-variable independent parameter $x(t)$,
provided that there is both a prompt ($<$ hrs) and a filtered ($>$ a
day) systemic response to the time variations in this parameter.
Numerical simulations of a version of this scenario where QPO
frequency depends on the balance between X-ray luminosity and
accretion rate via the accretion disk, and luminosity depends on this
accretion rate, both promptly and filtered by some averaging
mechanism, show a remarkable similarity to what is actually observed.

The basic prediction of this entire class of models is that it should
be possible to reconstruct QPO frequency from the time variations of
the governing parameter $x$ using a model involving some running
average of $x$.  In the example simulated, $x$ is identified with the
accretion rate through the disk $\dot M_d$ and $L_x$ is $\dot
M_d+\alpha\av{\dot M_d}$, where $\av{\dot M_d}$ is $\dot M_d$
convolved with a response function with characteristic time scale
$\tau$, so that we are faced with a deconvolution problem which might
in principle be solved using time series of simultaneous measurements
of instantaneous $L_x$ and QPO frequency long enough to cover a
sufficient number of correlation time scales $\tau$, and dense enough
to sample the short time scale variations well.  Perhaps this can be
accomplished by combining RXTE PCA and ASM observations, although PCA
observations extending over several weeks would be much preferable.

For models of the general type expressed in Eqs.\,1, both the
long-term and short-term variations in $L_x$ are caused by variations
in $\dot M_d$; the changes in kHz QPO frequency and presumably in the
other frequencies associated with the short-term $L_x$ variations
arise from changes in the inner disk radius $r$, which in turn are
related to changes in the ratio $\eta\equiv\dot M_d/\av{\dot M_d}$
between instantaneous and long-term average $\dot M_d$. Note, that
while the changes in X-ray spectral parameters and QPO amplitudes are
also affected by changes in this ratio, this occurs not necessarily
exclusively through the effect $\eta$ has on $r$. For the example
simulated, if both disk and radial accretion contribute to $L_x$ with
a different spectral shape, then the spectrum would be more similar if
the ratio of their rates is similar. If there are radiative-transfer
effects of the radial flow on spectrum or QPO amplitudes, these would
be stronger when the radial flow is denser; i.e., these effects depend
on $\av{\dot M_d}$ instead of on $\eta$.

So, in this scenario, the elusive time-variable parameter (often
dubbed ``inferred $\dot M$'' in the literature) that to first order
determines all QPO frequencies and a number of other timing and
spectral parameters (but not $L_x$) turns out to be $\eta$, the disk
accretion rate normalized by its own long-term average, rather than
any straight accretion rate. Curve length $S$ along the track of a
source in the color-color diagram is in this picture a measure of
$\eta$.  Depending on the precise nature of the filtered response
$\av{\dot M_d}$ (e.g., nuclear burning or radial accretion), $L_x$ may
or may not be a pure measure of the total accretion rate $\dot M$, but
even if it is, as in the model expressed in Eqs.\,2, neither $L_x$ nor
total $\dot M$ are directly correlated with $\eta$ and hence with all
the associated observable parameters.  Although the short-term
variations in $\eta$, $r$ and associated observable parameters such as
$S$ and $\nu$, as well as those in $L_x$ do indeed result directly
from short-term variations in the disk accretion rate $\dot M_d$, the
value of $\eta$ does not follow from the actual value of $\dot M_d$ or
$L_x$, nor from that of any other instantaneous accretion rate such as
total $\dot M$.

This leads to several further predictions.  Two observations
characterized by different $L_x$ but the same $\nu$ in this
description find the source with a disk that has the same $r$ and
hence $\eta$, but different $\dot M_d$ and $\av{\dot M_d}$. No strong
decrease of QPO fractional amplitude with luminosity is predicted as
would be the case if the luminosity variations would be dominated by
an extra source of X-rays unrelated to the QPOs --- this seems to be
borne out by observations (M\'endez et al. 2000). Which spectral
variations are seen for luminosity variations along or across parallel
lines depends on how variations in respectively $\eta$ and $\dot M$
affect the spectrum.  Motion from one parallel track to the other in
the $L_x$, $\nu$ diagram could happen along any trajectory depending
on the $\dot M_d$ record and the precise nature of the filtered
response, but no sudden jumps in the frequency-luminosity plane
connecting branches far apart would be expected. Note that {\it each}
correlated quantity is predicted to depend to first order on $\eta$;
two QPO frequencies, one depending on $\eta$ and another on $\dot M_d$
would not in the long run remain well correlated to each other.

The scenario discussed in this paper does not aim to explain the {\it
differences} with respect to spectral/timing properties observed as a
function of luminosity, which of course exist. Clearly, even if the
basic idea is correct that disk accretion rate normalized by its own
long-term average, $\eta=\dot M_d/\av{\dot M_d}$, to first order
determines most of the phenomenology either because it sets inner disk
radius $r$ or directly, it is not difficult to think of mechanisms
that could explain subtle differences in phenomenology at similar $r$
as a function of modest changes in $L_x$ such as described by, e.g.,
Kuulkers et al. (1996), and more obvious differences such as those
between Z and atoll sources for large $L_x$ differences (Hasinger and
van der Klis 1989). It is beyond the scope of this paper to speculate
on the precise nature of such mechanisms.

%dat er GX sources bestaan is wel een heel erg sterk argument voor
%Lx effecten natuurlijk}

\subsection{Black holes}

Finally, it is of interest to consider to what extent this type of
scenario could be of relevance to black-hole candidates. In those
sources, a similar type of correlation on short and decorrelation on
longer time scales is sometimes observed between $L_x$ on the one hand
and spectral hardness and timing parameters on the other (\S1). In the
black-hole candidate XTE J1550$-$564, parallel tracks are in fact
observed in an $L_x$ vs. spectral hardness diagram, with similar
hardness and QPO variations occurring along tracks in a hardness
vs. count rate diagram at very different luminosity (Homan et
al. 2001, who also discuss the possibility that $r$ varies more or
less independently from $L_x$). Perhaps a similar mechanism such as
that explaining the similar phenomenology in the neutron star systems
underlies this behavior, although matters are complicated by the
absence of a solid surface allowing (but not requiring) matter to
accrete without producing much radiation (e.g., Esin, McClintock and
Narayan 1997), so that the radiative stresses determining $r$ are
harder to predict from the accretion flows.

Reports of ``hysteresis'' in the flux from black-hole candidates
(Miyamoto et al. 1995, Smith, Heindl and Swank 2001; these latter
authors discuss their results in terms of two independent accretion
flows), with a tendency towards hard spectra when the count rate rises
and soft spectra when it falls, suggest that a disk plus radial flow
pattern related by a filtered response as described here could play a
role in these systems: if the count rate is dominated by the disk
accretion rate, and spectral hardness is representative of inner disk
radius then this is exactly what would be expected from the model
described in \S3: for a long-term dropping count rate $\eta>1$ so $r$
is small and the spectrum soft, and vv. This means that parallel
tracks could be produced in the way explained in this paper. 

Different from the neutron stars, where there are usually no
cross-tracks connecting the parallel tracks, the parallel tracks in
XTE J1550$-$564 are connected together in what was interpreted to be a
comb-like pattern (Homan et al. 2001). The very quiet, soft (``high'')
state the source is in when it is on the connecting track (the back of
the comb) may hence be unique to black holes, perhaps because it is
associated to a flow mode where the matter that accretes radially
flows into the hole without producing much radiation, so that the
radiative stress on the inner edge of the disk flow is low and hence
$r$ small and insensitive to small variations in the accretion
flows. The parallel tracks (the teeth of the comb) on which the other
black-hole states are observed (``very high'', ``intermediate'' and
``low'' states), may correspond to a mode where, perhaps due to
interaction with a jet, the radial flow does produce considerable
radiation so that $r$ is larger and sensitively dependent on the
accretion flows, as in neutron stars. This scenario would require the
radial flow to produce considerable radiation with low angular
momentum at radii smaller than $r$; it is unclear if there is a
mechanism by which this could be accomplished.

\section{Conclusion}

The scenario described in this paper can, for the first time, explain
how it is possible that kHz QPO frequency and several other timing and
spectral parameters are the same in sources differing by more than two
orders of magnitude in $L_x$, and in a given source for $L_x$ levels
different by a factor of several, even when those properties in any
given source at any given instant do themselves strongly depend on
$L_x$. The proposed solution is that such properties to first order
depend on $\eta$, disk accretion rate normalized by its own long-term
average. This at the same time removes the need for additional
independent, slowly varying or fixed source-dependent parameters
affecting the phenomenology, and provides a way in which most of the
observed short-term variations can arise as a consequence of
short-term accretion rate variations while preserving the expected
strong dependence of luminosity on total accretion rate.

The physical mechanism providing the long-term averaged response to
the disk accretion rate variations required in this scenario can be
the accretion of matter flowing off the disk vertically and accreting
radially, in combination with a mechanism to make the inner disk
radius dependent on the ratio of luminosity to disk accretion
rate. The radiation produced by this radial accretion flow in the
inner disk region is crucial in that case in determining much of the
phenomenology. If in black holes, contrary to neutron stars, there is
a mode for this radial flow to accrete while producing little
radiation (and another where it produces considerable radiation), then
this additional option available to black holes might explain the more
complicated phenomenology of those sources as compared to neutron
stars.

This work was supported in part by the Netherlands Organization for
Scientific Research (NWO). It is a pleasure to acknowledge
stimulating discussion and useful comments by Lars Bildsten, Mariano
M\'endez and Demetris Psaltis.

%\clearpage

\end{document}